\documentclass{article}

\usepackage{microtype}
\usepackage{graphicx}
\usepackage{subfigure}
\usepackage{booktabs} 
\usepackage{natbib} 
\usepackage{amssymb}
\usepackage{hyperref}
 \usepackage{amsmath}

\usepackage[accepted]{icml2019}

\begin{document}

\twocolumn[
\icmltitle{Orbit: Probabilistic Forecast with Exponential Smoothing}

\icmlsetsymbol{equal}{*}

\begin{icmlauthorlist}
\icmlauthor{Edwin Ng}{uber}
\icmlauthor{Zhishi Wang}{uber}
\icmlauthor{Huigang Chen}{uber}
\icmlauthor{Steve Yang}{uber}
\icmlauthor{Slawek Smyl}{uber}
\vskip 0.1in
$\mathtt{\{edwinng, ~zhishiw, ~huigang, ~steve.yang, ~slawek\}@uber.com}$

\end{icmlauthorlist}

\icmlaffiliation{uber}{Uber Technologies, Inc.}


\vskip 0.3in
]

\printAffiliationsAndNotice{} 


\begin{abstract}
Time series forecasting is an active research topic in academia as well as industry. Although we see an increasing amount of adoptions of machine learning methods in solving some of those forecasting challenges, statistical methods remain powerful while dealing with low granularity data. This paper introduces a  package \emph{Orbit} where it refined Bayesian exponential smoothing model with the help of probabilistic programming package such as Stan and Pyro. Our model refinements include additional global trend, transformation for multiplicative form, noise distribution and choice of priors. A benchmark study is conducted on a rich set of time-series data sets for our models along with other well-known time series models.
\end{abstract}


\section{Introduction}
\label{intro}

Time series forecasting is one of the most popular and yet the most challenging tasks, faced by researchers and practitioners. Its industrial applications have a wide range of areas such as social science, bioinformatics, finance, business operations and revenue forecast.  At Uber, time series forecasting has its applications from demand/supply prediction in the marketplace to the Ads budget optimization in the marketing data science.

In recent years, machine learning and deep learning methods \citep{gers1999learning, huang2015bidirectional, selvin2017stock} have gained increasing attention in the time series forecasting, due to their great ability in capturing the non-linear trend and complex interactions within multivariate time series. However, the theories for deep learning are still in the active research and development progress and not well established yet in the forecasting literature, especially in the case of univariate time series. At the same time, the blackbox nature of machine learning/deep learning models causes difficulties in interpretability and explainability \citep{gunning2017explainable, gilpin2018explaining}. In a benchmark study with different time series data \citep{hewamalage2019recurrent}, the authors show mixed performances of various deep learning models when compared against traditional statistical models.

On the other hand, traditional statistical parametric models have a well-established theoretical foundation, such as the popular autoregressive integrated moving average (ARIMA) \citep{box1968some} and exponential smoothing \citep{gardner1985exponential}. Recently, researchers from Facebook developed Prophet \citep{taylor2018forecasting}, which is based on an additive model where non-linear trends are fit with seasonality and holidays.

In this paper we propose a family of refined Bayesian exponential smoothing models, with great flexibility on the choice of priors, model type specifications, as well as noise distribution. Our model introduces a novel global trend term, which works well for short term time series. Most importantly, our models come with a well crafted compute software/package in Python, called Orbit (Object-oriented Bayesian Time Series). Our package leverages the probabilistic programming languages, Stan \citep{carpenter2017stan} and Pryo \citep{bingham2019pyro}, for the underlying MCMC sampling process and optimization. Pyro, developed by researchers at Uber,  is a universal probabilistic programming language (PPL) written in Python and supported by PyTorch and JAX on the backend. Orbit currently has a subset of the available prediction and sampling methods available for estimation using Pyro. 

The remainder of this paper is organized as follows. In \autoref{review}, a review is given on some popular statistical parametric time series models. In \autoref{orbit}, we give our refined model equations, and Orbit package design review. In \autoref{benchmark}, an extensive benchmark study is conducted to evaluate the proposed models’ performance and compare with other time series models discussed in the paper. \autoref{conclusion} is about the conclusion and future work. We focus on the univariate time series in this work.

\section{Review}
\label{review}

\subsection{Problem Definition}
Let $\{y_1, \cdots, y_t\}$ be a sequence of observations at time $t$. Then a point forecast at time $T$ denotes the process of estimating the set of future values of $\{\hat{y}_{T+1}, \cdots , \hat{y}_{T+h}\}$ given information at $t$ $\forall t\leq{T}$ where $h$ denotes the forecasting horizon.

\subsection{The State Space Models}
State space models is a family of models which can be written in a general form
$$y_t = Z_t^{T}\alpha_t + \epsilon_t$$
$$\alpha_{t+1} = T_t\alpha_t + R_t\eta_t$$

One big advantage is that they are modular, in the sense that independent state components can be combined by concatenating their observation vectors $Z_t$ and arranging the other model matrices as elements in a block diagonal matrix. This provides considerable flexibility for modeling trend, seasonality, regressors, and potentially other state components that may be necessary in practice.

However, such form is not unique since the same model can be expressed in multiple ways.  In practice, there are various reduced forms introduced by researchers.  We will go over some of those widely used in the industry.

\subsubsection{ARIMA}
ARMA model is one of the most commonly used methods to model univariate time series. ARMA($p$,$q$) combines two components: AR($p$), and MA($q$). In AR($p$) model, the value of a given time series, $y_t$, can be estimated using a linear combination of the $p$ past observations, together with an error term $\epsilon_t$ and a constant term $c$:
$$y_t = c + \Sigma_{i=1}^{p}\psi_i y_{t-i} + \epsilon_t$$
where $\psi_i, \forall i \in \{1,\cdots, p\}$ denote the model parameters, and $p$ represents the order of the model. Similarly, the MA(q) model uses past errors as explanatory variables:
$$y_t = \mu + \Sigma_{i=1}^{q}\theta_i \epsilon_{t-i} + \epsilon_t$$
where $\mu$ denotes the mean of the observations, $\theta_i, \forall i \in \{1,\cdots,q\}$ represents the parameters of the models and $q$ denotes the order of the model. Essentially, the method MA(q) models the time series according to the random errors that occurred in the past $q$ lags.

The model ARMA($p$,$q$) can be constructed by combining the AR($p$) model with the MA($q$) model, i.e.,
$$y_t = c + \Sigma_{i=1}^{p}\psi_i y_{t-i} + \Sigma_{i=1}^{q}\theta_i \epsilon_{t-i} +  \epsilon_t$$
The ARMA($p$,$q$) model is defined for stationary data. However, many realistic time series in practice exhibit a non-stationary structure, e.g., time series with trend and seasonality. The ARIMA($p$,$d$,$q$) (or SARIMA) overcomes this limitation by including an integration parameter of order $d$. In principle, ARIMA works by applying $d$ differencing transformations to the time series until it becomes stationary before applying ARMA($p$,$q$).

\subsubsection{Exponential Smoothing}

Another popular approach is done through exponential smoothing in a reduced form 

$$\hat{y}_{t|t-1} = Z^T \alpha_{t-1}$$
$$\epsilon_t = y_t - \hat{y}_{t|t-1}$$
$$\alpha_t = T\alpha_{t-1} + k\epsilon$$

This was first described by Box and Jekins \citep{box1998p}.  Note that $\hat{y}_{t|t-1}$ can be computed recursively. \citealt{hyndman2008forecasting} provide a complete review on such form.  In the literature, it introduces an $\textit{ETS}$ form with notation ``A" and ``M" which stands for additive and multiplicative respectively.  Well known methods such as Holt-Winter's model can be viewed as $\textit{ETS}(A, A, A)$ or $\textit{ETS}(A, M, A)$.  For notation such as $A_d$, the subscript $d$ denotes a damped factor is introduced.  Many of those are implemented under the \emph{forecast} package written in the statistical programming language R.

\subsubsection{Bayesian Structural Time Series (BSTS)}
Scott and Varian \citep{scott2014predicting} provide a bayesian approach on a reduced form as such

$$y_t = \mu_t + \tau_t + \beta^T{x_t} \epsilon_t$$
$$\mu_{t+1} = \mu_t + \delta_t + \eta_{0t}$$
$$\delta_{t+1}=\delta_t + \eta_{1t}$$
$$\tau_{t+1} = -\sum^{S-1}_{s=1} \tau_t + \eta_{2t}$$

This model is named as local linear trend model and is available in the \emph{bsts} package written in R. It allows additional regression estimation with coefficients $\beta$.

\subsection{Prophet}
Researchers at Facebook \citep{taylor2018forecasting} use a generalized additive model (GAM) with three main components: trend, seasonality, and holidays. They can be combined in the following equation:
$$y_t = g_t + s_t + h_t + \epsilon_t$$
where $g_t$ is the piecewise linear or logistic growth curve to model the non-periodic changes in the time series, $s_t$ is the seasonality term, $h_t$ is the holiday effect with irregular schedules, and $\epsilon_t$ is the error term. On a high level, Prophet is framing the forecasting problem as a curve-fitting exercise rather than looking explicitly at the time based dependence of each observation within a time series. As a computational tool/software, moreover, Prophet allows users to manually supply change points in fitting the trend term and set the boundaries for saturation growth, which gives great flexibility in business applications.

\section{The Orbit Package}
\label{orbit}

\subsection{Refined Model I - LGT}
Our proposed Local and Global Trend (LGT) model is an additive version based on the multiplicative model proposed in Rlgt \citep{smyl2019rlgt}.   In the forecast process, we have

\begin{align*}
 y_{t}  &= \mu_t + s_t   + \epsilon_t \\
\mu_t &= l_{t-1} + \xi_1 b_{t-1} + \xi_2 l_{t-1}^{\lambda}\\
\epsilon_t  &~\sim \mathtt{Student}(\nu, 0, \sigma)\\
\sigma &~\sim  \mathtt{HalfCauchy}(0, \gamma_0)
\end{align*}

where $l_{t-1}, \xi_1 b_{t-1}, \xi_2 \l_{t-1}^{\lambda}, s_{t}$ and ${\epsilon}_t$ can be viewed as level, local trend, global trend, seasonality and error term, respectively. 

In the update process, it is similar to the triple exponential smoothing form
\begin{align*}
l_t &= \rho_l(y_t - s_t) + (1-\rho_l)l_{t-1}\\
b_t &=  \rho_b(l_t - l_{t-1}) + (1-\rho_b)b_{t-1}\\
s_{t+m} &=  \rho_s(y_t - l_t ) + (1-\rho_s)s_t
\end{align*}

where $\rho_l$, $\rho_b$ and $\rho_s$ are the smoothing parameters.  Finally, we fit $\rho_l$, $\rho_b$, $\rho_s$,  $\theta$, $\xi$, $\lambda$, $\nu$ in our training process while $\gamma_0$ is a data-driven scaler.

This model structure is similar to the $\textit{ETS}$ family established by Rob Hynd \citep{de2011forecasting, hyndman2018forecasting} except that the form we proposed is a hybrid model generalizing both $\textit{ETS}(A,A,A)$ (when $\xi=0$) and autoregressive model (when $\xi > 0$).
A similar approach is found earlier in (Smyl, Zhang et al., 2015).   The advantages over Rlgt with such form are two-fold.  First, the computation is more effective on the additive form, where we can apply simple log transformation to maintain multiplicative properties.  Second, $\sigma$ in Rlgt is parameterized with $y_t$ which introduces dependency in noise generation, while $\sigma$ is refined as an independent noise in LGT. This change reduces the computation cost by vectorizing the noise generation process.

One limitation in this model is that it assumes $l_t > 0$,  $\forall t$.  To ensure such condition is satisfied during the training period, it imposes a requirement such that $y_t >0$, $\forall t$.

\subsection{Refined Model II - DLT}

For use cases with $y_t \in \mathbb{R}$, we provide an alternative - Damped Local Trend (DLT) model. 

Such model is an extension of $\textit{ETS}(A,A_d,A)$ in \citealt{hyndman2008forecasting}, where  $\theta$ is known as the damped factor.  In the forecast process, we have

\begin{align*}
y_t &=\mu_t + s_t + r_t +  \epsilon_t \\
\mu_t &=D(t) + l_{t-1} +  \theta{b_{t-1}}
\end{align*}

with the update process as such 

\begin{align*}
g_t &= D(t)\\
l_t &= \rho_l(y_t - g_{t} - s_t - r_t) + (1-\rho_l)(l_{t-1} + b_{t-1})\\
b_t &=  \rho_b(l_t - l_{t-1}) + (1-\rho_b)\theta{b_{t-1}}\\
s_{t+m} &=  \rho_s(y_t - l_t - r_t) + (1-\rho_s)s_t\\
r_t &=  \Sigma_{j}\beta_j x_{jt}
\end{align*}

There are a few choices of $D(t)$ as a deterministic global trend.  Options such as  \emph{linear},  \emph{log-linear} and  \emph{logistic} are included in the package.  Another feature of DLT is the introduction of regression component $r_t$. This serves the purpose of nowcasting or forecasting when exogenous regressors are known such as events and holidays.  Without loss of generality, assume

$$\beta_j ~\sim \mathtt{Normal}(\mu_j, \sigma_j)$$

where $\mu_j = 0$  and $\sigma_j = 1$ by default as a non-informative prior. There are more choices of priors for the regression component in the package.

\subsection{Multiplicative Form}

In previous section, we derive a general additive form for LGT and DLT. It is trivial that by using transformation $y^{\prime} _t =  ln(y_t)$, our forecast process yields a multiplicative form as such
$$\hat{y}_t= \mu^\prime_t \cdot s^\prime_t \cdot r^\prime_t$$

In the later benchmarking process,  we used models fitted in such multiplicative form to report performance metrics.

\subsection{Package Design}
Orbit is our Python package to implement the refined models discussed above. This package is written and designed from a strict object oriented perspective with the goals of re-usability, ease of maintenance, and high efficiency.

\begin{figure}[ht]
\vskip 0.2in
\begin{center}
\centerline{\includegraphics[width=\columnwidth]{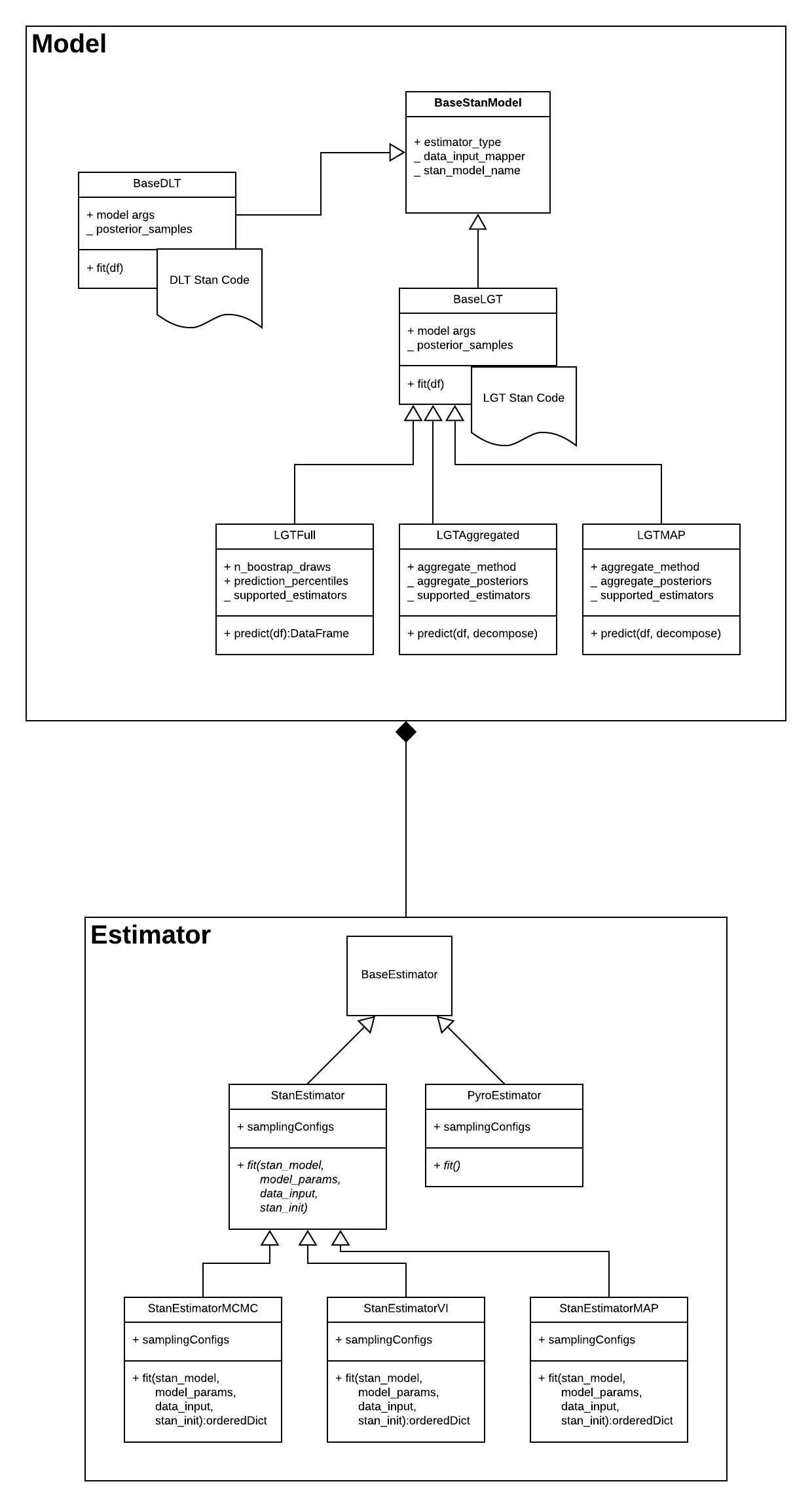}}
\caption{Overall Design of Orbit Package.}
\label{design}
\end{center}
\vskip -0.2in
\end{figure}

The base Estimator class contains generic logic to handle interaction with the underlying inference engine (e.g PyStan, Pyro) along with utilities to load and save Orbit models, and the specifics of the model are implemented in each model class. \autoref{design} is the overall workflow of Orbit package.

\section{Model Benchmark}
\label{benchmark}

\subsection{Data}
We performed a comprehensive benchmark study on five datasets:
\begin{itemize}
\item US and Canada rider first-trips with Uber (20 weekly series by city)
\item US and Canada driver weekly first-trips with Uber (20 weekly series by city)
\item Worldwide first-orders with Uber Eats(15 daily series by country)
\item M3 series (1428 monthly series)
\item M4 series (359 weekly series)

\end{itemize}

where M3/M4 time series are well-known in the forecast community \citep{makridakis2018m4}.

\subsection{Performance Metric}
We use symmetric mean absolute percentage error (SMAPE), a widely adopted forecast metric, as our performance benchmark metric

$$\mathtt{SMAPE} = \sum^{h}_{t=1} \frac{|F_t - A_t|}{(|F_t| + |A_t|)/2} $$
where $X_t$ represents the value measured at time $t$ and $h$ is the forecast horizon.

Here $h$ is the forecast horizon which can also be considered as the ``holdouts" in a backtest process.  Following what competitions suggested, we use 13 forecast horizon and 18 forecast horizon, respectively, for M4 weekly and M3 monthly series with 1 split; for Uber datasets, we use $h=13$, 3 splits with 26 incremental steps for weekly series and $h=28$, 4 splits and 14 incremental steps for daily series. With multiple splits, we expect more robust result. The calculation is done with the help of our backtest utilities built in the Orbit package.

\subsection{Results}

We compared our proposed models, LGT and DLT, to other popular time series models such as SARIMA \citep{seabold2010statsmodels} and Facebook Prophet \citep{taylor2018forecasting}. Both Prophet and Orbit models use Maximum A Posterior (MAP) estimates and they are configured as similar as possible in terms of optimization and seasonality settings. For SARIMA, we fit the $(1,1,1) \times {(1, 0, 0, )}_S$ structure by maximum likelihood estimation (MLE) where $S$ represents the choice of seasonality. 

Within a dataset, the models in consideration were run on each time series separately, and then we report the aggregated metrics. \autoref{map-table} gives the average and the standard deviation (within parentheses) of SMAPE across different models and datasets.


\begin{table}[t]
\caption{Model Average SMAPE Comparison}
\label{map-table}
\begin{center}
\begin{tiny}
\begin{sc}
\begin{tabular}{lcccr} \hline

Model & LGT & DLT & Prophet & SARIMA \\ \hline

Rider-weekly & 0.108 (0.030) & {\bf 0.106} (0.033)& 0.121 (0.035) &0.110 (0.041) \\

Driver-weekly & {\bf 0.205} (0.098) & 0.206 (0.095) & 0.274 (0.162) & 0.207 (0.123)\\

Eater-daily & {\bf 0.178} (0.036) & 0.182 (0.043) & 0.309 (0.055) & 0.221 (0.040)\\

M4-weekly & 0.077 (0.073)&  0.081 (0.080)&  0.192 (0.321) & {\bf 0.076} (0.131)\\ 

M3-monthly & {\bf 0.145} (0.155) & 0.148 (0.158) & 0.193 (0.234) &0.150 (0.171)\\ \hline

\end{tabular}
\end{sc}
\end{tiny}
\end{center}
\vskip -0.1in
\end{table}

%

It shows that our models consistently deliver better accuracy than other candidate time series models in terms of SMAPE. Orbit is also computationally efficient. For example, the average compute time per series with full MCMC sampling and prediction from a subset of M4 weekly data is about 2.5 minutes and 16 ms. The run time for the same series in Prophet is about 10 minutes for sampling and 2.4 s for prediction. That’s a 4x speed up in training, and orders of magnitude difference in prediction.

Code and M3/M4 data used in this benchmark study are available upon request.

\section{Conclusion}
\label{conclusion}
We have shown that our proposed models outperform the baseline time series models consistently in terms of SMAPE metrics. Furthermore, we also identified compute cost improvements when using the Orbit package. For our future work, we will continue to actively maintain the package, incorporate new models, and provide new features or enhancements (to support dual seasonality, fully Pyro integration, etc).


\nocite{hyndman2018forecast}

\bibliography{orbit_paper}

\begin{thebibliography}{20}
\providecommand{\natexlab}[1]{#1}
\providecommand{\url}[1]{\texttt{#1}}
\expandafter\ifx\csname urlstyle\endcsname\relax
  \providecommand{\doi}[1]{doi: #1}\else
  \providecommand{\doi}{doi: \begingroup \urlstyle{rm}\Url}\fi

\bibitem[Bingham et~al.(2019)Bingham, Chen, Jankowiak, Obermeyer, Pradhan,
  Karaletsos, Singh, Szerlip, Horsfall, and Goodman]{bingham2019pyro}
Bingham, E., Chen, J.~P., Jankowiak, M., Obermeyer, F., Pradhan, N.,
  Karaletsos, T., Singh, R., Szerlip, P., Horsfall, P., and Goodman, N.~D.
\newblock Pyro: Deep universal probabilistic programming.
\newblock \emph{The Journal of Machine Learning Research}, 20\penalty0
  (1):\penalty0 973--978, 2019.

\bibitem[Box(1998)]{box1998p}
Box, G.
\newblock \emph{P, Jenkins, GM and Reinsel, GC. Time series analysis:
  forecasting and control}, volume~54, pp.\  176--180.
\newblock 1998.

\bibitem[Box \& Jenkins(1968)Box and Jenkins]{box1968some}
Box, G.~E. and Jenkins, G.~M.
\newblock Some recent advances in forecasting and control.
\newblock \emph{Journal of the Royal Statistical Society. Series C (Applied
  Statistics)}, 17\penalty0 (2):\penalty0 91--109, 1968.

\bibitem[Carpenter et~al.(2017)Carpenter, Gelman, Hoffman, Lee, Goodrich,
  Betancourt, Brubaker, Guo, Li, and Riddell]{carpenter2017stan}
Carpenter, B., Gelman, A., Hoffman, M.~D., Lee, D., Goodrich, B., Betancourt,
  M., Brubaker, M., Guo, J., Li, P., and Riddell, A.
\newblock Stan: A probabilistic programming language.
\newblock \emph{Journal of statistical software}, 76\penalty0 (1), 2017.

\bibitem[De~Livera et~al.(2011)De~Livera, Hyndman, and
  Snyder]{de2011forecasting}
De~Livera, A.~M., Hyndman, R.~J., and Snyder, R.~D.
\newblock Forecasting time series with complex seasonal patterns using
  exponential smoothing.
\newblock \emph{Journal of the American statistical association}, 106\penalty0
  (496):\penalty0 1513--1527, 2011.

\bibitem[Gardner~Jr(1985)]{gardner1985exponential}
Gardner~Jr, E.~S.
\newblock Exponential smoothing: The state of the art.
\newblock \emph{Journal of forecasting}, 4\penalty0 (1):\penalty0 1--28, 1985.

\bibitem[Gers et~al.(1999)Gers, Schmidhuber, and Cummins]{gers1999learning}
Gers, F.~A., Schmidhuber, J., and Cummins, F.
\newblock Learning to forget: Continual prediction with lstm.
\newblock 1999.

\bibitem[Gilpin et~al.(2018)Gilpin, Bau, Yuan, Bajwa, Specter, and
  Kagal]{gilpin2018explaining}
Gilpin, L.~H., Bau, D., Yuan, B.~Z., Bajwa, A., Specter, M., and Kagal, L.
\newblock Explaining explanations: An overview of interpretability of machine
  learning.
\newblock In \emph{2018 IEEE 5th International Conference on data science and
  advanced analytics (DSAA)}, pp.\  80--89. IEEE, 2018.

\bibitem[Gunning(2017)]{gunning2017explainable}
Gunning, D.
\newblock Explainable artificial intelligence (xai).
\newblock \emph{Defense Advanced Research Projects Agency (DARPA), nd Web}, 2,
  2017.

\bibitem[Hewamalage et~al.(2019)Hewamalage, Bergmeir, and
  Bandara]{hewamalage2019recurrent}
Hewamalage, H., Bergmeir, C., and Bandara, K.
\newblock Recurrent neural networks for time series forecasting: Current status
  and future directions.
\newblock \emph{arXiv preprint arXiv:1909.00590}, 2019.

\bibitem[Huang et~al.(2015)Huang, Xu, and Yu]{huang2015bidirectional}
Huang, Z., Xu, W., and Yu, K.
\newblock Bidirectional lstm-crf models for sequence tagging.
\newblock \emph{arXiv preprint arXiv:1508.01991}, 2015.

\bibitem[Hyndman et~al.(2008)Hyndman, Koehler, Ord, and
  Snyder]{hyndman2008forecasting}
Hyndman, R., Koehler, A.~B., Ord, J.~K., and Snyder, R.~D.
\newblock \emph{Forecasting with exponential smoothing: the state space
  approach}.
\newblock Springer Science \& Business Media, 2008.

\bibitem[Hyndman \& Athanasopoulos(2018)Hyndman and
  Athanasopoulos]{hyndman2018forecasting}
Hyndman, R.~J. and Athanasopoulos, G.
\newblock \emph{Forecasting: principles and practice}.
\newblock OTexts, 2018.

\bibitem[Hyndman et~al.(2018)Hyndman, Athanasopoulos, Bergmeir, Caceres, Chhay,
  O'Hara-Wild, Petropoulos, Razbash, Wang, and Yasmeen]{hyndman2018forecast}
Hyndman, R.~J., Athanasopoulos, G., Bergmeir, C., Caceres, G., Chhay, L.,
  O'Hara-Wild, M., Petropoulos, F., Razbash, S., Wang, E., and Yasmeen, F.
\newblock forecast: Forecasting functions for time series and linear models.
\newblock 2018.

\bibitem[Makridakis et~al.(2018)Makridakis, Spiliotis, and
  Assimakopoulos]{makridakis2018m4}
Makridakis, S., Spiliotis, E., and Assimakopoulos, V.
\newblock The m4 competition: Results, findings, conclusion and way forward.
\newblock \emph{International Journal of Forecasting}, 34\penalty0
  (4):\penalty0 802--808, 2018.

\bibitem[Scott \& Varian(2014)Scott and Varian]{scott2014predicting}
Scott, S.~L. and Varian, H.~R.
\newblock Predicting the present with bayesian structural time series.
\newblock \emph{International Journal of Mathematical Modelling and Numerical
  Optimisation}, 5\penalty0 (1-2):\penalty0 4--23, 2014.

\bibitem[Seabold \& Perktold(2010)Seabold and Perktold]{seabold2010statsmodels}
Seabold, S. and Perktold, J.
\newblock statsmodels: Econometric and statistical modeling with python.
\newblock In \emph{9th Python in Science Conference}, 2010.
\newblock Package version 0.11.1.

\bibitem[Selvin et~al.(2017)Selvin, Vinayakumar, Gopalakrishnan, Menon, and
  Soman]{selvin2017stock}
Selvin, S., Vinayakumar, R., Gopalakrishnan, E., Menon, V.~K., and Soman, K.
\newblock Stock price prediction using lstm, rnn and cnn-sliding window model.
\newblock In \emph{2017 international conference on advances in computing,
  communications and informatics (icacci)}, pp.\  1643--1647. IEEE, 2017.

\bibitem[Smyl et~al.(2019)Smyl, Bergmeir, Wibowo, and Ng]{smyl2019rlgt}
Smyl, S., Bergmeir, C., Wibowo, E., and Ng, T.~W.
\newblock Bayesian exponential smoothing models with trend modifications.
\newblock \emph{https://cran.r-project.org/web/packages/Rlgt/index.html}, 2019.

\bibitem[Taylor \& Letham(2018)Taylor and Letham]{taylor2018forecasting}
Taylor, S.~J. and Letham, B.
\newblock Forecasting at scale.
\newblock \emph{The American Statistician}, 72\penalty0 (1):\penalty0 37--45,
  2018.
\newblock Package version 0.7.1.

\end{thebibliography}
\bibliographystyle{icml2019}

\end{document}